\documentclass[reprint,
amsmath,amssymb,
aps,
]{revtex4-2}

\newcounter{schemes}
\usepackage{graphicx}\usepackage{dcolumn}\usepackage{bm}\usepackage{hyperref}\usepackage{verbatim}

\usepackage{xcolor}
\usepackage{mathtools}

\newcommand{\fig}[1]{Fig. (\ref{#1})}
\newcommand{\eq}[1]{Eq. (\ref{#1})}

\newcommand{\ie}{\emph{i.e.}}
\newcommand{\etal}{\emph{et al.}}
\newcommand{\tb}{time-bin}
\newcommand{\tbs}{time-bins}
\newcommand{\rt}{round-trip}
\newcommand{\rts}{round-trips }
\newcommand{\tr}{\tau_r}
\newcommand{\tpos}{\Delta\tau}

\begin{document}

\preprint{APS/123-QED}
\title{Demonstration of a quantum C-NOT Gate in a Time-Multiplexed fully reconfigurable photonic processor}\author{Federico Pegoraro$^\dagger$}
\email{federico.pegoraro@uni-paderborn.de}
\affiliation{Paderborn University, Integrated Quantum Optics, Institute of Photonic Quantum Systems (PhoQS) Warburger Str. 100, 33098, Paderborn, Germany}
\author{Philip Held$^\dagger$}
\affiliation{Paderborn University, Integrated Quantum Optics, Institute of Photonic Quantum Systems (PhoQS) Warburger Str. 100, 33098, Paderborn, Germany}
\author{Jonas Lammers$^\dagger$}
\affiliation{Paderborn University, Integrated Quantum Optics, Institute of Photonic Quantum Systems (PhoQS) Warburger Str. 100, 33098, Paderborn, Germany}
\author{Benjamin Brecht}
\affiliation{Paderborn University, Integrated Quantum Optics, Institute of Photonic Quantum Systems (PhoQS) Warburger Str. 100, 33098, Paderborn, Germany}
\author{Christine Silberhorn}
\affiliation{Paderborn University, Integrated Quantum Optics, Institute of Photonic Quantum Systems (PhoQS) Warburger Str. 100, 33098, Paderborn, Germany}
\date{\today}

\begin{abstract}
The two-qubit controlled-not (C-NOT) gate is an essential component for gate-based quantum circuits. In fact, its operation, combined with single qubit rotations allows to realise any quantum circuit.
Several strategies have been adopted in order to build quantum gates.
Among them, photonics offers the dual advantage of excellent isolation from the environment and ease of manipulation at the single qubit level.
Here we adopt a scalable time-multiplexed approach in order to build a fully reconfigurable architecture capable of implementing a post-selected C-NOT gate with a fidelity of $(93.8\pm1.4)\%$. We then show how our time-multiplexed platform can be employed to combine a C-NOT and a single qubit gate in order to generate the four Bell states.
\\
\begin{description}
\item[\large$\dagger$]These authors contributed equally
\end{description}
\end{abstract}

\maketitle
The past decades witnessed a momentous race towards advances in quantum computation and communication science.
At the foundation of any quantum computation scheme lays the quantum bit, or qubit: the fundamental information unit consisting of a quantum system with two levels corresponding to the logical $|0\rangle$ and $|1\rangle$ states.
A variety of physical platforms have been employed in order to implement systems of one or multiple qubits: notable ones employ for instance the spin of charged particles in semiconductors \cite{chargedparticlesSpinQubit}, atomic nuclei \cite{NMRquantumComputer}, trapped ions \cite{trappedIONqcumputer} and quantum dots \cite{quantumDotsComputer}; others exploit vibrational energy levels in molecules \cite{vibrationalQubits}, superconductive systems \cite{supercondictiveQubits}, or topological properties of two-dimensional systems \cite{topologicalQuantumComputer}.\\
One of the possible paradigms is photonic quantum computing (PQC) \cite{knill2001scheme,kok2007linear,PhysRevA.64.062311}.
In PQC photons are the carriers of quantum information, which is encoded in the degrees of freedom of the light quanta.
This approach offers both excellent isolation between system and external environment and ease of manipulation of the single qubit states via linear optical elements combined with measurement-induced non-linearities.\\
Complementary to information encoding is the need for operations capable of manipulating the information content of the quantum system of choice.
For this reason it is necessary to be able to build logic quantum gates capable of acting on the state of one or many qubits.
While single qubit gates can alter the state of individual information units, two- or more qubit operations influence the joint state of a system and introduce interaction between subsystems.
Among the two-qubit gates the controlled-not (C-NOT) gate stands out for its capability of generating entanglement and is an essential component for the construction of arbitrary gate-based quantum circuits \cite{gateBasedQuantumComputing}.
In spite of the inherently non-interactive nature of photons, the realization of a C-NOT gate based on linear optics is rendered possible by the adoption of ancillary modes and coincidence measurements in different interferometric schemes \cite{ralph2002linear,PhysRevA.64.062311}.
Experimental implementations have demonstrated the C-NOT adopting several encoding strategies involving a variety of light properties: many exploited path and polarization encoding both in bulk \cite{o2003demonstration,weipanBulkNonDestructiveCNOT} and integrated \cite{crespi2011integrated,femtosecondWGCNOT,wahlterHeraldedIntegratedCnot,zhang2021supercompact,he2023super} optics, in other instances degrees of freedom such as frequency \cite{frequencyBinCNOT,POLFREQCNOTlu2024building} or orbital angular momentum of light \cite{OAMCNOT} have been employed.\\
For scaling systems time-multiplexing (TM) has been explored in recent years \cite{QWREFschreiber2010photons,QWREFschreiber2011decoherence,QWREFdynamicnitsche2016quantum,QWREFpegoraro2023dynamic} and has actually been harnessed to achieve quantum computation advantages in Gaussian boson sampling \cite{madsen2022quantum}.
Yet, these systems are intrinsically not capable of achieving universal quantum computation.\\ 
Possible time-encoded schemes capable of implementing single- and two-qubit gates have already been demonstrated by Lo \emph{et al.} in \cite{lo2020quantum} and by Humphreys \emph{et al.} in \cite{timeMultiplexedCPHASE}.
In both cases the light temporal degree of freedom has been used to demonstrate a post-selected C-PHASE gate. 
Here we report on the first direct demonstration of a photonic C-NOT gate in a TM experimental platform featuring fast electro-optical modulators which can be programmed to implement an optical interferometer in the time domain.
While in \cite{lo2020quantum} time is used to encode the qubit's degree of freedom, the gate relies on separated spatial inputs for the two qubits. 
In contrast, our experimental platform, similarly the one in \cite{timeMultiplexedCPHASE}, requires only a single spatial mode where qubits are multiplexed on a series of temporal bins.
This allows for implementing a circuit with many input modes and, in principle, many input qubits.
At the same time, compared to the platform shown in \cite{timeMultiplexedCPHASE}, ours is fully and rapidly reconfigurable and readily suited to implement feed-forward.\\   
Using our system, we show a TM circuit capable of implementing the scheme of \cite{ralph2002linear}, and we provide a characterization of its behavior.
We use the platform to reconstruct the C-NOT logical table of truth and obtain an excellent agreement with the expected behavior.
Lastly, we show that our system can be programmed to implement both the C-NOT and single qubit operations.
By combining these two primitive elements, we implement a quantum circuit capable of generating the four Bell states $|\Phi^{\pm}\rangle$ and $|\Psi^{\pm}\rangle$.
This highlights both the entangling power of our C-NOT, and how our TM approach constitutes a suitable and reconfigurable hardware for the implementation of single qubit rotations as well as two qubit interactions, both essential requirements for the realization a universal quantum computer.    

\section{Results}

\subsection{Time-multiplexed interferometric scheme} \label{sub:RES_TMscheme}

\begin{figure}[!h]
\centering
   \includegraphics[width=.7\columnwidth]{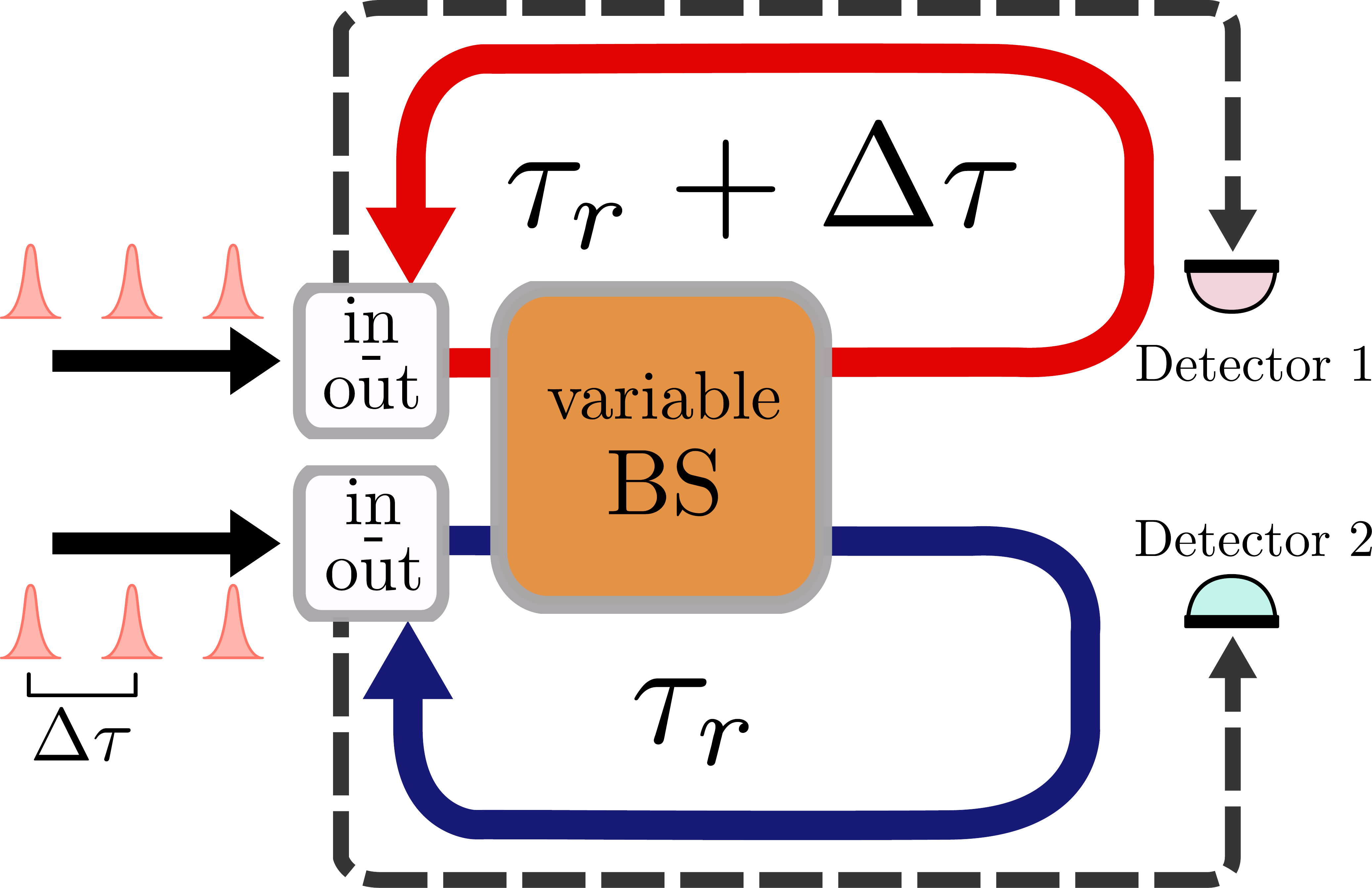}
   \caption{ 
   Conceptual scheme of the TM interferometer.  
   }
   \label{fig:concept}
\end{figure}
Our interferometric approach is based on the implementation of a unitary transformation between input and output TM modes.
These are encoded by a sequence of \tbs$\ $separated by a time $\tpos$.\\ 
In order to implement a transformation on consecutive \tbs, we consider the scheme shown in Fig. \ref{fig:concept}, where we can identify a structure featuring a short and a long delay loop with travel times $\tr$ and $\tr+\tpos$.    
The two loops are connected by a variable beam splitter (BS) whose reflectivity may be set independently for any \tb.
For a parametrization of this element in terms of a 2$\times$2 unitary see the Methods.\\ 
Photons distributed on a stream of \tbs$\ $may enter or exit the looped architecture thanks to the action of two devices located within the loops (white in-out boxes in Fig. \ref{fig:concept}).\\ 
A single \rt$\ $in this interferometer allows to interfere neighboring \tbs.
\begin{figure}[!h]
\centering
   \includegraphics[width=.9\columnwidth]{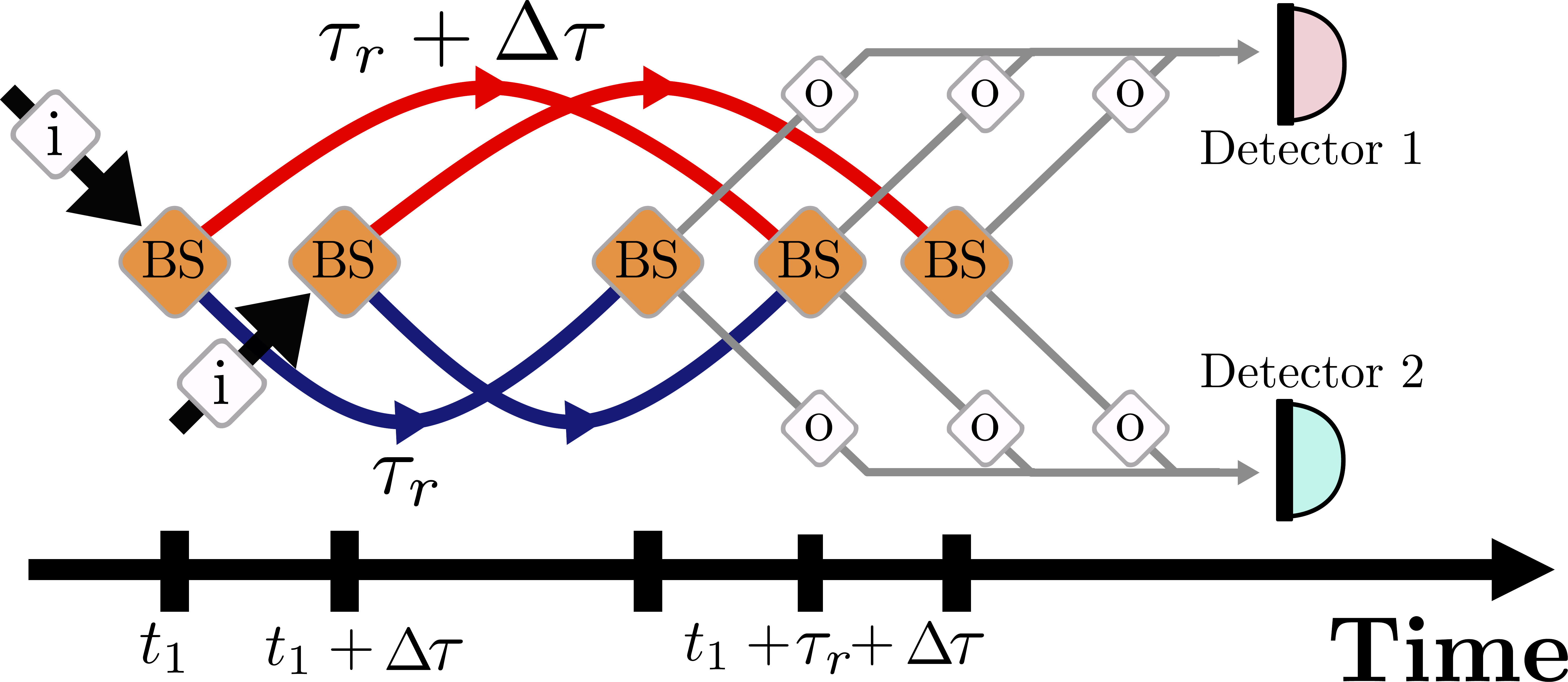}
   \caption{  
   Timeline for a single \rt$\ $in the loop. Two consecutive \tbs$\ $. 
   Interference among the two \tbs$\ $ is possible considering that time $t_1+\tr+\tpos$ may be reached by both travelling either in the long or short loop.
   Performing $n$ \rts$\ $it is possible to interfere \tbs$\ $spaced by $n\tpos$. 
   }
   \label{fig:timeline}
\end{figure}
This is shown in Fig. \ref{fig:timeline}, where we consider two \tbs$\ $which are in-coupled and reach the variable BS a times $t_1$ and $t_1+\tpos$.
Photons travelling the short loop (blue) reach the next BS after a time $\tr$, while those travelling the long loop (red) after a time $\tr+\tpos$.
By equalling the time difference between loops and \tbs$\ $we ensure that photons, that were initially separated by $\tpos$, can be recombined one $\rt$$\ $at $t=t_1+\tau_r+\tpos$, where they interfere when returning to the central BS.

Following this principle, \tbs$\ $spaced by $n\tpos$ can be interfered performing $n$ \rts$\ $in the interferometer.
A readout of the output is possible using a detector for each loop capable of resolving the output \tb$\ $stream.\\         
\subsection{Experimental setup}
\begin{figure*}
   \includegraphics[width=.9\textwidth]{setup.pdf}
   \caption{TM setup: photons at 1545 nm with H and V polarization enter the setup and are directed to the short and long single mode fiber, respectively. After travelling though the fibers they have accumulated a relative delay of 170 ns. 
   EOM 1 and EOM 2 flip the photon polarizations in order to redirect both to the feedback where EOM 3 acts on their polarization states before travelling again to the fibers.
   At each  \rt$\ $EOM 3 can implement an arbitrary series of operations wich realize the optical interferometer.
   After the desired amount of \rts$\ $EOM 1 and 2 act again on the photon polarization to send the output \tb$\ $stream to the detection unit described in the Methods.}
   \label{fig:setup}
\end{figure*}
We realise this scheme using the setup shown in Fig. \ref{fig:setup}.
In our implementation, the variable BS of Fig. \ref{fig:concept} is an electro-optical modulator (EOM) capable of implementing an arbitrary polarization rotation (EOM 3 in Fig. \ref{fig:setup}). 
Thus, our interferometer acts both on the temporal and polarization degrees of freedom of the propagating photons. 
The setup input is located at a polarization BS (PBS), PBS 1 in the setup sketch, where H- and V-polarized photons are separated and sent to two single mode fibers using fiber collimators.
The fibers serve as delay loops and have lengths of 1085 and 1120 m.
In order to maintain the photon polarization state we use a set of waveplates positioned before the input collimators.\\ 
At the fiber outputs, two fiber collimators equal to the input ones, send H and V photons to two EOMs (EOM 1 and 2).
These are capable of rotating their polarization state by $90^{\circ}$, and are used as in- and out-coupling units. 
A second PBS (PBS 2) directs photons to EOM 3 where the BS operation can take place on the desired \tbs.
All EOMs are base on bulk Pockels cells featuring an optical transmission of $\geq$ 99\%. 
A detailed discussion of the operation of the three EOMs can be found in the Methods.\\
The temporal separation between successive \tbs$\ \tpos$ is set to $170$ ns, while the \rt$\ $time is $\tr=5.3$ $\mu$s. 
$\tpos$ is chosen to be larger than the minimum pulse separation for EOM 3, which is the slowest of the EOMs, therefore, each \tb$\ $can be addressed independently.\\
Instead, the \rt$\ $time allows our system to operate on up to $30$ \tbs, which combined with the polarization encoding for the variable BS, means that our system features a total of $60$ possible input and output modes.
The number of modes is only limited by $\tr$ and can be increased using longer fibers. 
In this setup we achieve an average fiber to fiber coupling efficiency of $81\%$.
This value consists of the average of four different fiber coupling, namely the two self-coupling efficiencies from a loop to itself and the two cross-coupling form a loop to the other. 
To ensure that the four loop efficiencies are as similar as possible the setup is built symmetrically, \ie$\ $the distances between input and output fiber collimators and their focal lengths are equal.
This ensures that the spatial mode propagating in the free space section of the setup is the same for both delay loops. 
Additionally, the loops are iteratively aligned using the flexure stages and mirror mounts shown in the setup sketch in order to equalize all couplings, trough this procedure we obtain individual efficiencies differing no more than $1\%$ with respect to each other.
When the input state occupies only a single time-bin, interference occurs only if the photons have traversed the short and long loops an equal number of times. Consequently, interfering photons accumulate the same phase and experience identical dispersion in both fibers.
This renders the setup robust against temperature-induced phase fluctuations and effects of fibre dispersion.\\  
As input, we use signal-idler photon pairs produced using type-II spontaneous parametric down-conversion (PDC) in a 2.5 cm long periodically poled potassium-tytanil phospate (ppKTP).  
The process is pumped at generating a mean photon number of $\langle n\rangle=0.01$ wich sets the pair generation probability to 1\% and limits the influence of higher order photon number contributions.
The waveguide exhibits losses of 1.2 dB/cm, and since we can approximate the generation as occurring at its midpoint, each photon exits the waveguide with a probability of 70\%.
The two photons are spectrally degenerate with a central wavelength of $1545$ nm, they have a temporal duration of approx. $3$ ps, are temporally synchronized, and feature an indistinguishability of $(98\pm1)\%$.
Since signal and idler are produced by a type-II process, they are in an horizontal and vertical polarization state, respectively.
More information about the source setup and the optimization of photon idistinguishability can be found in the supplementary information.\\
After performing the desired number of \rts$\ $in the TM interferometer, photons are out-coupled using EOMs 1 and 2 and directed to a detection setup comprising four superconducting nanowire single-photon detectors (SNSPDs) with a dead time of approximately 70 ns, timing jitter of 120 ps, and efficiencies $\geq 90\%$.
Although the SNSPDs exhibit dark counts of up to 100 cps, their impact is mitigated through time gating with a 5 ns window centered on each respective time-bin.
This limits the probability of registering a dark count in a system output to only 0.005\% per bin. 
The SNSPDs are integrated into a setup that enables time- and polarization-resolved detection, as well as quantum state tomography, the details of which are provided in the Methods.
Overall, when performing a single pass through the TM setup, the system features an efficiency of 21\% from source to detection.    

\subsection{Demonstration of the TM C-NOT gate}\label{sub:CNOT_res}

\begin{figure}
    \centering
    \includegraphics[width=0.70\columnwidth]{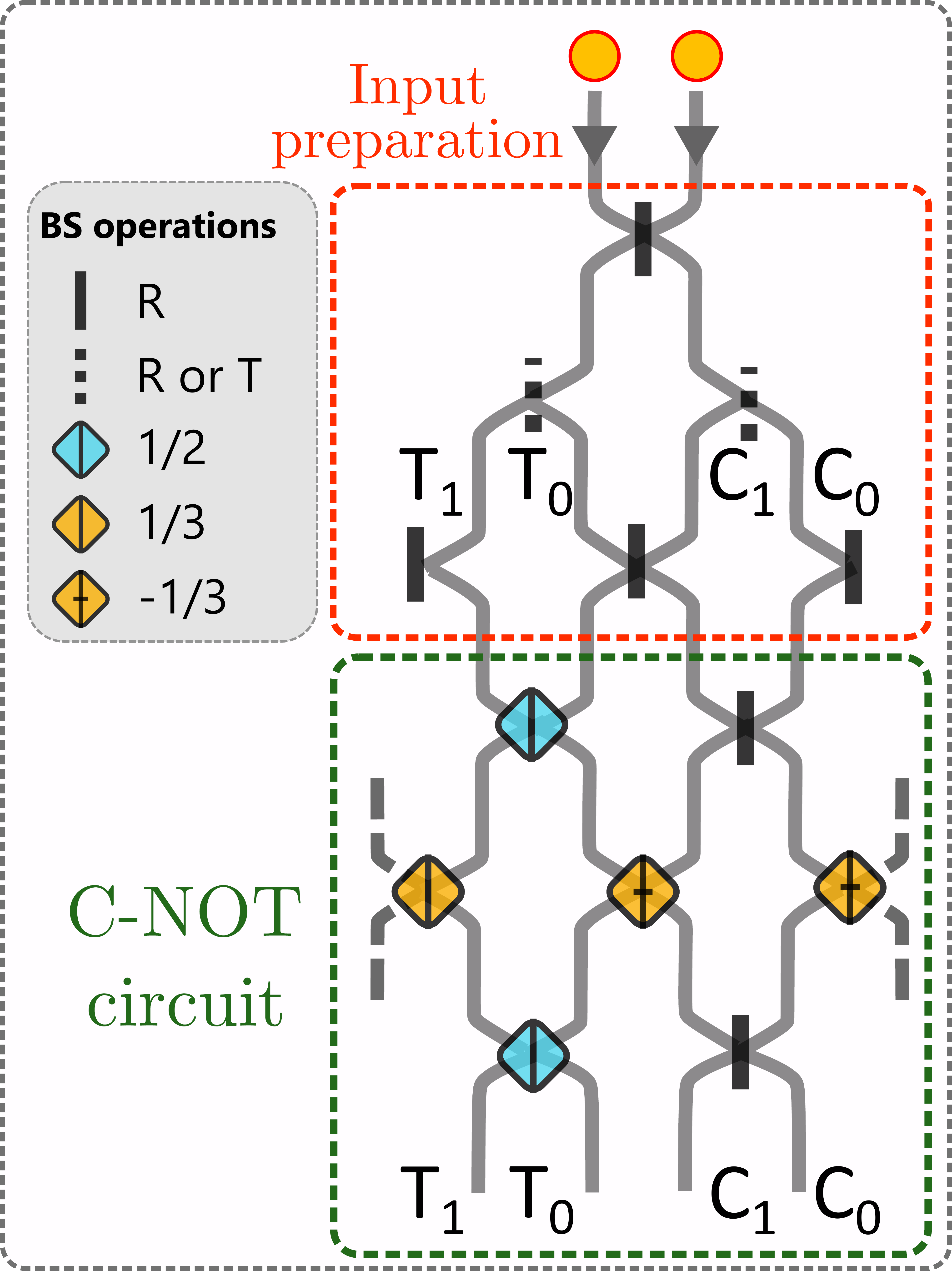}
    \caption{Path encoded circuit implementing both state preparation and C-NOT gate.
    In the first block of the interferometer a series of reflections and transmissions is used to direct one photon to the control ($C_0$,$C_1$) and one to the target ($T_0$,$T_1$) modes.
    The next section implements the gate and it is analogous to the interferometer proposed in \cite{ralph2002linear}.
    The gate is successful upon detecting a photon in the control and one in the target outputs. 
    The dashed paths correspond to ancillary modes required to implement the gate.}
    \label{fig:pathcircuit}
\end{figure}

The C-NOT gate is an operation involving a control (C) and a target (T) qubit.
Similarly to what happens in its classical counterpart, the state of the T-qubit is flipped if and only if the C-qubit is in the logical $|1\rangle$ state.
The action of the C-NOT is represented by a unitary transformation, which in the C-T computational basis \{$|00\rangle$,$|01\rangle$,$|10\rangle$,$|11\rangle$\} takes the form:
\begin{equation}
    \mathcal{U}_{C-NOT}=
    \begin{pmatrix}
        1 &0 &0 &0\\
        0 &1 &0 &0\\
        0 &0 &0 &1\\
        0 &0 &1 &0\\
    \end{pmatrix}.
    \label{eq:UC-NOT}
\end{equation}
The main challenge in the realization of a photonic C-NOT gate resides in the inherently non-interacting nature of photons. 
Interaction can be introduced by coupling photons to ancillary modes and exploiting post-selection on specific coincidence patterns \cite{ralph2002linear,wahlterHeraldedIntegratedCnot}, as well as two particle interference \cite{hong1987measurement}. 
In the previous section we have expanded on how we implement an optical interferometer in a TM fashion.
However, before describing TM C-NOT circuit, we will consider the path encoded system shown in Fig. \ref{fig:pathcircuit} which corresponds to the one that will be implemented in time. 
The optical network consists of 6 layers of BSs, which can be grouped into an input state preparation section and a stage implementing the C-NOT.
 At the first layer we have a single BS performing a full reflection, two photons enter from the two BS inputs and are sent to the next layer where the dashed lines indicate two BSs performing either a reflection or a transmission.
 The actual operation depends on which of the four possible input states we generate.
 As shown in the figure, the two modes on the left side encode the T-qubit, while the remaining modes are devoted to the C-qubit.
 Therefore, the four C-T possible inputs are generated according to the following BS operations:
 \begin{equation}
 \begin{split}
	&C_0\ T_0\rightarrow R\ T\\
	&C_0\ T_1\rightarrow T\ T\\
	&C_1\ T_0\rightarrow R\ R\\
	&C_1\ T_1\rightarrow T\ R,
\end{split}
	\label{eq:inputgen}
\end{equation}      
where $R$ and $T$ denote a reflection and transmission, respectively. 
The third layer of the network features only reflections and therefore only redirects the photons.\\
The next three layers implement the C-NOT interferometer of Ralph \etal \cite{ralph2002linear}, where reflections in the C modes ensure that the $|0\rangle_c$ and $|1\rangle_c$ states are kept separated.
Therefore, $C_0$ will only interact with one of the ancillary modes (represented by the dashed line) through a BS with reflectivity $-\frac{1}{3}$. 
The remaining part of the circuit implements a Mach-Zehnder interferometer between $T_0$ and $T_1$ whose arms are coupled on the left to another ancillary mode and to the right with $C_1$ by means of a $\frac{1}{3}$ and $-\frac{1}{3}$ BS, respectively.\\    
At the end, target and control are distributed on four outputs and the gate is successful upon post selecting events involving coincidences among modes: $(C_0,T_0)$, $(C_0,T_1)$, $(C_1,T_0)$, $(C_1,T_1)$, which in this example are output paths, while in the TM implementation are encoded by a combination of \tbs$\ $and polarization.\\ 
In order to translate this into the TM scheme we consider the sequence of operations shown in Fig. \ref{fig:timecircuit}.
Here, the various layers of the circuit shown in Fig. \ref{fig:pathcircuit} are translated to \rts$\ $within the TM architecture.   
The photon pair starts in the same \tb$\ t=0$ with opposite polarizations.
In our setup, this corresponds to the arrival of the photon pair to PBS 1 (see Fig. \ref{fig:setup}), which implements the equivalent of the path encoded reflection by sending H and V to the short and long fiber, respectively. 
\begin{figure*}
    \centering
    \includegraphics[width=0.99\textwidth]{circuit_time.pdf}
    \caption{Temporally encoded version if the circuit shown in \fig{fig:pathcircuit} implemented by the setup shown in Fig. (\ref{fig:concept}). 
    Here, the action the many path beam splitters is translated into a series of operations carried out by a single device at specific times after the photons enter the setup ($t=0$).
    Blue and red arrows correspond respectively to traveling in the short and long loop, while the background color indicates polarization of light traveling in the loop.
    The polarization flips occurring at first and last \rt$\ $ corresponds to the action of the in- and out-coupling EOMs (EOM 1 and 2 in Fig. (\ref{fig:setup})).
    The gate outcome is read out by means of time- and polarization-resolved detection. 
    }
    \label{fig:timecircuit}
\end{figure*}
The in-coupling EOMs (EOM 1 and 2) act before the next sequence of operations, therefore, as shown in Fig. \ref{fig:timecircuit}, the polarizations traveling within short and long loop are flipped with respect to the input ones. 
All the remaining operations on the timeline are performed by EOM 3 and the circuit is performed by traveling a total of six \rts$\ $in the setup.
In order to perform all round-trips before a new experimental run is initiated and to ensure optimal operation of all EOMs, we run the experiment at a repetition rate of $\nu_{exp}=$21 kHz.
Taking into account photon generation probability, setup total efficiency and exponential losses in the loop, we can accumulate a total of 4000 two-photons event per hour of integration after six round-trips.
This combined with the C-NOT success rate, results in $\approx$450 post-selected events contributing to successful instances of the gate.\\          
After out-coupling every \tb, they are sent outside the setup, where the circuit's output is accessed performing polarization- and time-resolved detection.\\ 
The four input states are generated acting with EOM 3 according to Eq. (\ref{eq:inputgen}), additional details about the voltage sequences needed for this purpose can be found in the Methods.\\       

\begin{figure}
    \centering
    \includegraphics[width=0.99\columnwidth]{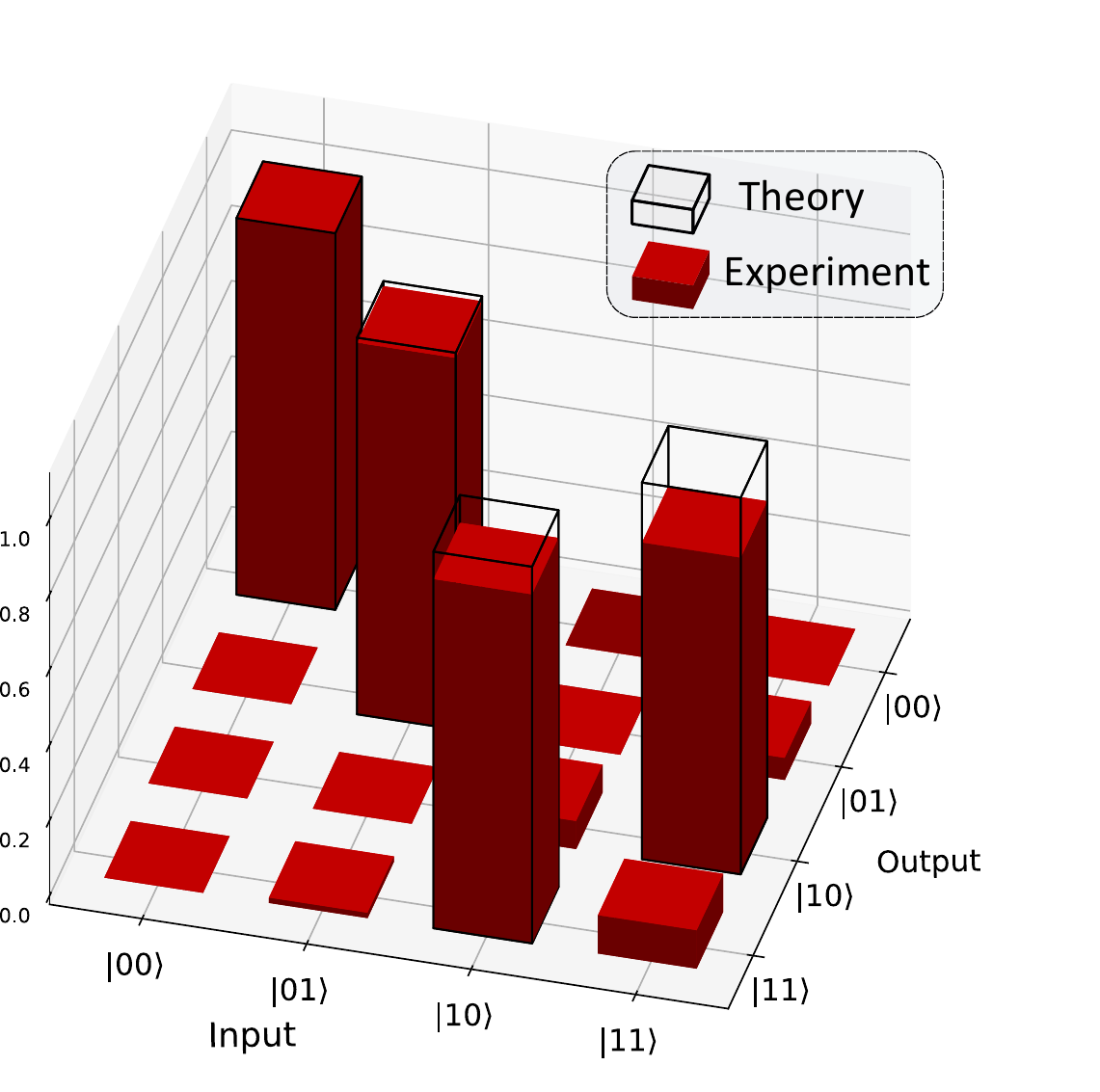}
    \caption{Measured vs theoretical truth table for the TM C-NOT. 
    From the measured coincidences we obtain operation probabilities of probability $P_{00}=(100.0\pm0.2)\%$ and $P_{01}=(98.6\pm0.2)\%$ for the $|00\rangle$ and $|01\rangle$ inputs, $P_{11}=(92.6\pm0.2)\%$ and $P_{10}=(84.1\pm0.4)\%$, for the $|10\rangle$ and $|11\rangle$ inputs. We find a gate operation fidelity of $(93.8\pm1.4)\%$. }
    \label{fig:truthtable}
\end{figure}

In order to validate the gate action we record time-stamps relative to the C- and T- output qubits, which, as shown in Fig \ref{fig:timecircuit}, are encoded in a combination of three \tbs.
Although $C_1$ and $T_0$ occupy the same \tb, we can still resolve them as they are in different polarization states.\\  
From the time-stamps we then reconstruct the coincidences between control and target and obtain the operation pattern shown in Fig. \ref{fig:truthtable}, where the red bars represent the normalized coincidences between control and target at the gate output for each input state. 
The experimentally reconstructed output probabilities are shown in comparison with the one expected from the C-NOT unitary (see \eq{eq:UC-NOT}).    
From this comparison we obtain a fidelity of $(93.8\pm1.4)\%$ with respect to the ideal truth table. 
The obtained gate fidelity compare favorably with implementations of the same post-selected interferometric scheme both in bulk \cite{o2003demonstration} and integrated \cite{crespi2011integrated,zhang2021supercompact,he2023super} optics.
Two main limiting factors bounding the operation fidelity are the residual distinguishability of the two photons and spurious multi-photon generation events from the SPDC process.
In the supplementary material we provide a detailed discussion of these effects. 
Our analysis shows that, even if the gate operation were perfect, the two aforementioned effects would limit the measured fidelity to $\leq95\%$. 
Comparing this to our measured fidelity of 93.8\%, we find that our circuit's performance is close to perfect.

\begin{figure}[h!]
    \centering
    \includegraphics[width=\columnwidth]{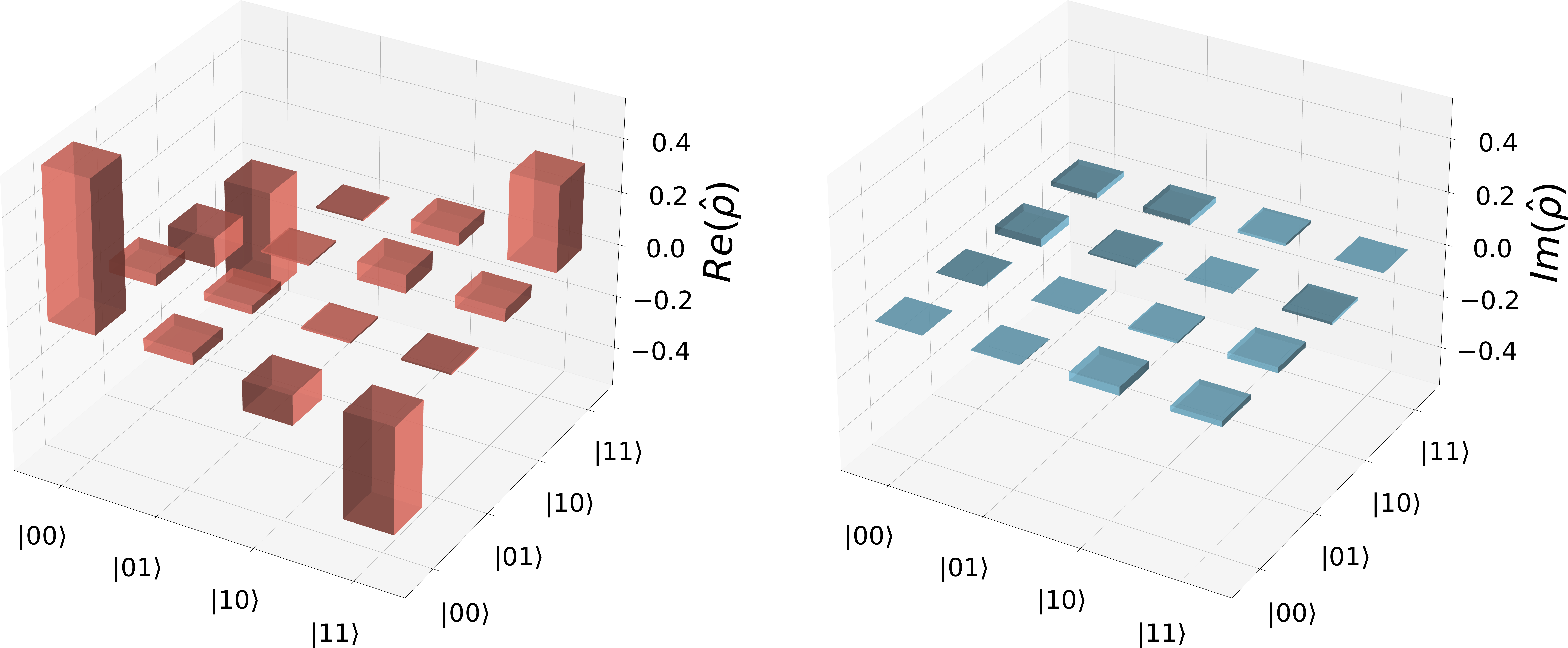}
    \caption{
Real and imaginary part of the reconstructed density operator for the $|\Phi^{-}\rangle$ state for which we observe a quantum fidelity $(85.0\pm2.4)\%$.}
    \label{fig:BELL}
\end{figure}

After verifying that the TM circuit is capable of implementing the gate, we use it to generate the four Bell states.
This requires to initiate the control qubit in an even superposition of $|0\rangle$ and $|1\rangle$.
This is accomplished programming EOM 3 to implement a $\frac{1}{2}$ switch on both \tbs$\ $at the first \rt$\ $of the C-NOT circuit (green box in Fig. \ref{fig:timecircuit}).
In addition to that, we perform an extra \rt$\ $where a series of reflections are used to separate target and control, which is required to perform a tomography of the resulting two-qubit state.\\ 
A complete scheme of the TM circuit used to generate the four Bell states can be found in the Methods.
Considering the series of operations required, our circuit maps the two-qubit computational basis in the following way:
\begin{align*}
    |00\rangle_{CT}\longrightarrow|\Phi^{-}\rangle_{CT}\\
    |01\rangle_{CT}\longrightarrow|\Psi^{-}\rangle_{CT}\\
    |10\rangle_{CT}\longrightarrow|\Phi^{+}\rangle_{CT}\\
    |11\rangle_{CT}\longrightarrow|\Psi^{+}\rangle_{CT}
\end{align*}
Using the tomography setup described in the Methods we performed a reconstruction of the density operators, for which we observe quantum state fidelities with respect to the target states of $\mathcal{F}_{\Psi^{-}}=(78.1\pm2.8)\%$, $\mathcal{F}_{\Psi^{+}}=(85.6\pm3.0)\%$, $\mathcal{F}_{\Phi^{-}}=(85.0\pm2.4)\%$ and $\mathcal{F}_{\Phi^{+}}=(80.3\pm3.0)\%$. 
In Fig. \ref{fig:BELL} we show real and imaginary part of the reconstructed density operator for the state $|\Phi^{-}\rangle$, the plots for the remaining ones can be found in the supplementary material.
 \section{Discussion}
In this work, after discussing our TM scheme, we have shown a platform that exploits dynamical polarization operations to translate the photonic C-NOT scheme from a path to a temporal encoding with high fidelity.
Moreover, the employed experimental platform grants us the ability to deterministically prepare different input control-target states, including superpositions at the single qubit level. 
This feature allows us to exploit our source efficiently by using both photons from a single generation event.\\
Additionally, using the fact that we can reconfigure the active elements of our system, we have shown how it can be employed to combine single- and two-qubit gates to generate quantum entanglement.\\
The system would also support, states involving more than two photons which may be implemented populating more than a single input \tb.\\
However, transmission efficiency within the TM setup is a critical parameter.

%In future experiments improving the mode matching, it would be reasonable to achieve a loop efficiency of 85\%, which would produce an increase of roughly 60\% in the amount of two-fold events after the two-photon gate.
In future experiments, improved mode matching would reasonably allow for achieving a loop efficiency of 85\%, which yields a roughly 60\% increase of two-fold events after the two-photon gate.
Further improvements could be obtained by reducing the fiber length, which would become feasible with faster EOMs.
In the present implementation, the EOM responsible for the beam-splitting operations has a bandwidth limited to 6.7 MHz, resulting in a time-bin separation of 170 ns.
The availability of comparable devices operating at speeds of up to 100 MHz in the foreseeable future would reduce the required time-bin separation by approximately one order of magnitude.
This would provide the dual benefit of increasing loop efficiencies to 90\% and enhance the achievable experimental data rates by a factor of thirty.
A detailed analysis of the achievable rates and their possible improvement can be found in the supplementary information.\\
Nevertheless, the TM interferometer offers a clear scalability advantage. 
In fact, large networks can, in principle, be realized by adjusting the fiber lengths while keeping the number of active components constant. 
This contrasts with other prominent photonic platforms, such as path-encoded systems, where both circuit size and number of required components (e.g., directional couplers and phase shifters) grows with the number of modes, increasing fabrication complexity, susceptibility to imperfections, and crosstalk. 
Moreover, such circuits must typically be fully programmed prior to operation, and partial intermediate readout of selected modes is not straightforward.
In contrast, the in- and out-coupling EOMs in our TM architecture can be programmed to address specific subsets of time bins.
These subsets can be analyzed by fast external logic, which can then dynamically determine subsequent operations for all EOMs.\\ 
Taken together, these features establish the TM architecture as a scalable, reconfigurable, and feed-forward–compatible platform that provides a compelling pathway toward the realization of large-scale photonic quantum circuits.

 \section{Acknowledgements}
The authors thank René Pollmann for the fruitful discussions. 
We acknowledge financial support by the European Commission through the Horizon Europe project EPIQUE
(Grant No. 101135288).
 \section{Author contributions}
F.P., P.H., J.L. performed the experiments and analyzed the results.
P.H. conceived the experimental scheme.
F.P. performed the state reconstruction and led writing of the manuscript.
P.H., J.L. assisted in writing the paper. 
C.S. and B.B. supervised the work.
C.S. initiated the work.
All authors discussed the results and commented on the manuscript.
\pagebreak \clearpage{}\section{Methods}

\subsection{Operation of electro optical modulators}
The TM architecture shown in the main text heavily relies on using electro optical modulators to act on specific time-bins. 
In particular, we employ EOMs for two specific purposes within the network:
\begin{itemize}
\item To deterministically in- and out-couple photons, 
\item To act on the polarization state of time-bins inside the loop in order to realise the desired circuit.
\end{itemize}
EOMs 1 and 2 (see Fig. \ref{fig:setup}) are devoted to the former task, while EOM 3 carries out the latter. 
Although devices performing the two different tasks feature different timings, speeds and operation modes, all modulators exploit Pockels effect \cite{pockels1889ueber}. 
Therefore, the operation of both kind of EOMs relies on a series of electrical signals with an appropriate voltage level and timing applied to their Pockels cells to alter the light polarization at the desired times.\\  
Temporal stability of the driving signals is ensured deriving the experimental triggers and clocks directly from the cavity of the laser used to pump the photon source.         

\subsubsection{Input/output switches}
\begin{figure}[h!]
\centering
    \includegraphics[width=.99\columnwidth]{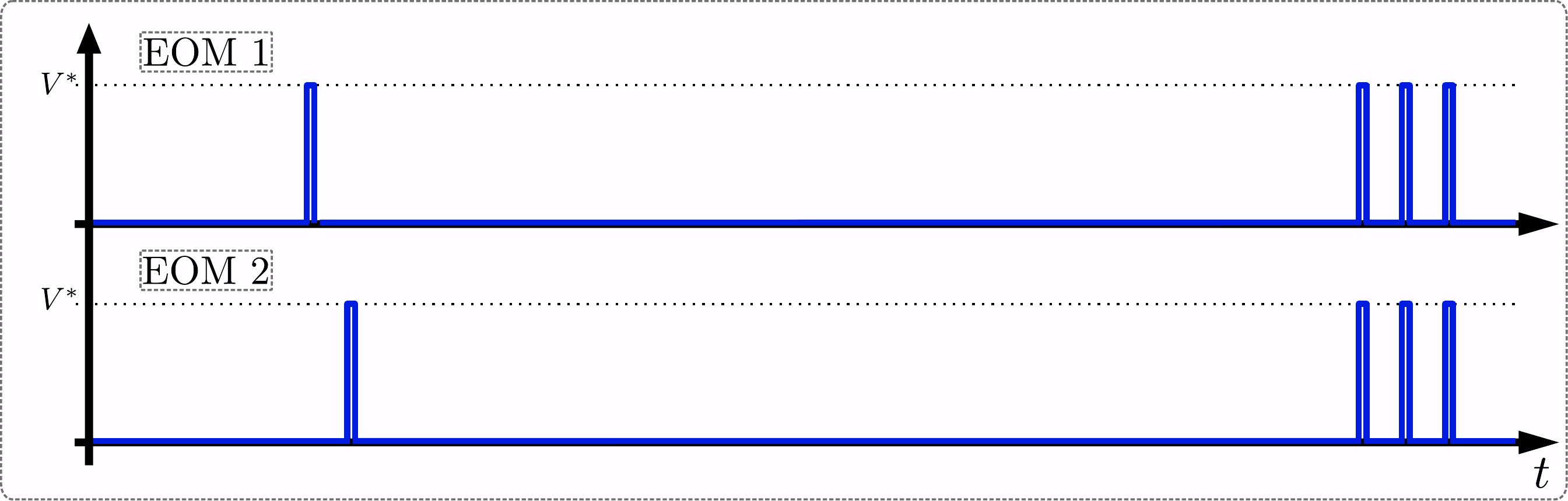}
    \caption{In- and out- coupling switching sequence for EOMs 1 and 2.}\label{fig:eom12switches}
\end{figure}
 
As stated in the main text input and output are implemented by EOMs 1 and 2 in our setup. 
Their Pockels cells, consiststing of Rubidium Tytanil-phosphate crystals (RTP), are adjusted to avoid clipping the impinging beam and affecting its polarization when no voltage is applied.\\
The circuit driving the cell can either be off or provide a voltage $V^*$, whose value is set by a potentiometer present on the EOM high voltage supply and is equal for all activations of the device.
We use these EOMs to perform a full polarization swap, which allows us to either direct photons to the setup feedback or outside in order to perform detection.\\
In order to generate the driving signals, we use delay generator cards provided by the vendor.
To operate, the delay generator cards receive a trigger signal whose frequency corresponds to the experimental repetition rate and a clock signal at $76.4$ MHz.  
With this we generate switching sequences that look like the ones shown in Fig. \ref{fig:eom12switches}. 
The two sequences start with a single switch corresponding to the arrival of the two photons to EOMs 1 and 2 after traveling through the fibers.
Since EOM 2 is located after the longer fiber its action is delayed with respect to EOM 1 exactly by $170$ ns. 
After these two switches, the photons remain inside the loop unless we flip again their polarization.
This happens after $5$ additional \rts, after which a sequence of three switches send the photons outside the setup. 
The three switches are spaced by $170$ ns and correspond to three output time-bins shown in Fig. \ref{fig:timecircuit}.
Given the fact that the minimum pulse spacing for these EOMs is $50$ ns, we can act independently on each \tb.
Although not required in our case this allows to use this system to partially in- and out-couple light and has been used to probe measurement induced effects \cite{nitsche2018probing} in previous version of the setup.

\subsubsection{Arbitrary polarization rotation}
\begin{figure}[h!]
\centering
    \includegraphics[width=.99\columnwidth]{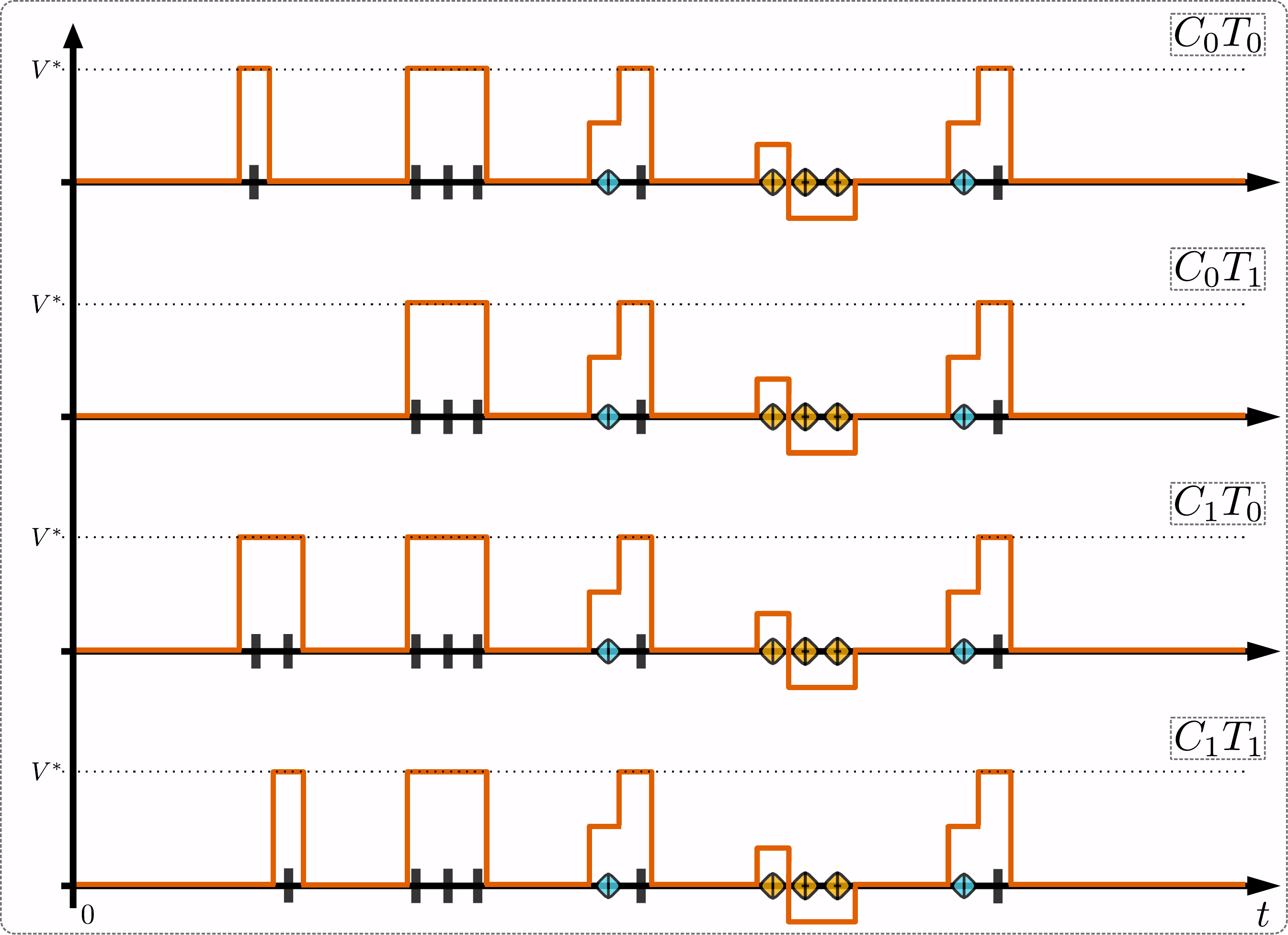}
    \caption{ Switching sequences for EOM 3. The applied voltage in the first round-trip  depends on what input C-T state is sent to the C-NOT, the remaining sequence realises the actual circuit and is set to match the BS reflectivities and phases of the scheme from Ralph $et\ al.$ \cite{ralph2002linear}.}\label{fig:eom4switches}
\end{figure}
The remaining EOM (EOM 3 in Fig. \ref{fig:setup}) is used to implement arbitrary polarization operations within the loop. 
Similarly to EOMs 1 and 2, the cell is made of RTP and can be aligned to the beam by means of an opto-mechanical mount provided by the vendor. 
In contrast to the other EOMs the voltage applied to the cell can take a different value for each switch. 
For this reason this EOM is driven by an arbitrary signal generator that receives the experimental trigger and a $10$ MHz clock. 
The minimum time between two switches is specified to be $150\ ns$, because of this parameter the time-bin separation is set to $170\ ns$ which allows the EOM to act independently on any time-bin.\\
Fig. \ref{fig:eom4switches} shows the switching sequences applied to EOM 3 by means of the arbitrary signal generator for the four different C-T input states. 
Overlaid to the time-line, we show the same BS symbols used in Fig. \ref{fig:timecircuit} to show the correspondence between voltage and circuital operation.
The reflectivity $R$ of each BS is determined by the voltage applied to the Pockels cell $V$ according to the relation: $R=sin^2(\frac{\pi}{2V^*}V)$, where $V^*$ is the voltage that causes a $90^{\circ}$ rotation of the polarization state.\\
Following this relation, full  reflections correspond to $V=V^*$, while operations involving $\frac{1}{2}$ and $\pm\frac{1}{3}$ BSs are implemented by the voltages $V=\frac{V^*}{2}$ and $V=\pm0.39V^*$, respectively.\\  
From the four switching sequences we see that only the first group of switches changes for the four different inputs: this is the case since the four possible input states for control and target correspond to different combinations of time-bins and polarizations. 
The remaining voltage levels are the same for all patterns and correspond to a redirection operated by the three reflective switches and the C-NOT circuit that takes place over the remaining three \rts$\!$.
To check that the BS operations are correctly implemented, we record a histogram of the system output triggered by the same signal used to initiate the experiment.
When a horizontally polarized CW laser is injected into the setup, the resulting histogram exhibits a series of dips corresponding to the activations of EOM 3.
This allows us to verify and optimize each splitting ratio by evaluating the contrast of each dip, defined as the ratio between the number of counts inside and outside the dip.
Temporal synchronization is achieved by applying a global delay to the operation sequence so that the first switching event matches the corresponding time bin.
Additionally, we ensure that the relative timing between successive dips agrees with the expected values determined by $\tau_r$ and $\Delta\tau$.\\
\subsection{Modelling of beam splitters and effect of phase shifts}
The BS constitutes the core element of many linear optical networks. Its action on the state of an incoming photon is parametrized by the unitary operation:
\begin{equation}\label{eq:BS_generalunitary}
    \hat{U}(R,T,\phi_{R},\phi_{T})=
    \begin{pmatrix}
    \sqrt{R}e^{i\phi_{R}} & \sqrt{T}e^{-i\phi_{T}}\\
    \sqrt{T}e^{i\phi_{T}} & -\sqrt{R}e^{-i\phi_{R}}
    \end{pmatrix},
\end{equation}
where $R$ and $T$ can be physically understood as the BS reflectivity and transmissivity, which in the lossless case satisfy the energy conservation condition $R+T=1$. The two remaining parameters $\phi_{R}$ and $\phi_{T}$ represent the phases imposed on transmitted and reflected light, respectively. In this work we consider BS operations where $\phi_T=0$ and reflection phases $\phi_R$ of either $0$ or $\pi$. Taking into account the relation between transmission and reflection parameters, the class of BS unitaries that we want to implement reduce to:    
\begin{equation}\label{eq:waveplate_BS}
    \hat{U}_{\pm R}=
    \begin{pmatrix}
        \pm \sqrt{R} & \sqrt{1-R}\\
        \sqrt{1-R} & \mp \sqrt{R}
    \end{pmatrix}.
\end{equation}
We restrict to these two particular case because it allows to model all BSs present in the scheme of \cite{ralph2002linear}. 

\subsection{TM Bell state generation}
\begin{figure*}
    \centering
    \includegraphics[width=.99\textwidth]{bell_methods.pdf}
    \caption{
        a) Bell state generation circuit
        b) TM circuit that generates Bell states.
   A 50:50 splitting implements the Hadamard gate required to put the control qubit in a superposition.
   Three reflections at the end ensure that target and control are in two separated \tbs.}
    \label{fig:bells_methods}
\end{figure*}
Two qubits starting in one of the four states: $|00\rangle$, $|01\rangle$, $|10\rangle$ and $|11\rangle$ may be transformed into the Bell states applying the circuit shown in Fig. \ref{fig:bells_methods}a). 
The circuit consists of an Hadamard and a C-NOT gate: the former acts on the first qubit to put its state in a superposition, while the latter acts on the joint state according to the unitary shown in Eq. (\ref{eq:UC-NOT}).
With a slight modification of the TM circuit we used to implement the C-NOT gate (see Fig. \ref{fig:timecircuit}) we can implement the Hadamard gate required to act on the control qubit.
This is shown in Fig. \ref{fig:bells_methods}b), where the input state generation stage is the same as the one shown in the main text and allows to generate the two qubit computational basis states starting from photon pairs produced by the soruce.
The main differences are in the remaining part of the circuit, where now also the later \tb$\ $in the first \rt$\ $of this stage implements a 50:50 splitting.
This operation implements the Hadamard gate applied to the control qubit and does not interfere with the C-NOT as it happens before control and target are mixed.        
Compared to the TM C-NOT, this circuit features an extra \rt$\ $at the end, where only reflections are applied. 
This is not strictly required to implement the circuit  of Fig. \ref{fig:bells_methods}a), however, it is necessary in order to reconstruct the states using the tomography setup as it requires the two photons encoding for the two qubits to be in a defined \tb.
This would not be the case without the redirection step. 
In fact, from Fig. \ref{fig:timecircuit} it is possible to see that both control and target are distributed onto two \tbs. 
Although the four possible outcomes can still be distinguished using polarization, the fact that $C_1$ and $T_0$ occupy the same \tb$\ $would render impossible to apply the reconstruction unitaries independently on the two qubits.
With the additional the redirection \rt$\ $we overcome this as target and control become temporally separated.

\subsection{Two-Qubit tomography setup}
\begin{figure}[h!]
\centering
    \includegraphics[width=.9\columnwidth]{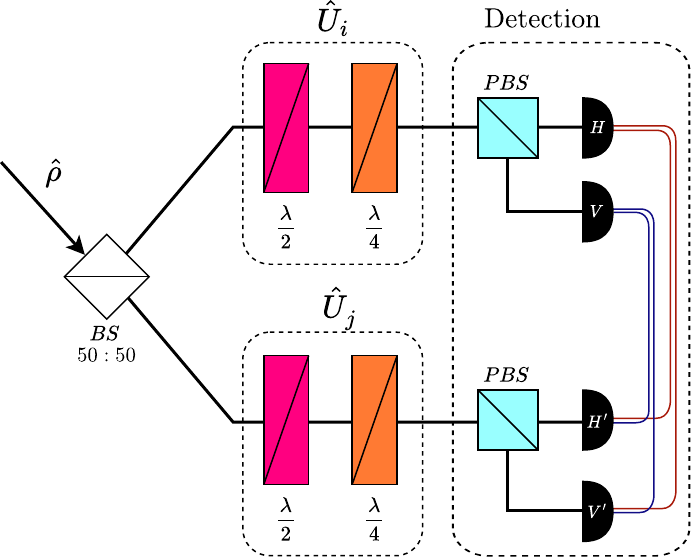}
    \caption{Sketch of the tomography setup. 
    The 50:50 BS splits photons probabilistically and send them to two units consisting of an half- and quarter-waveplate that rotate the state into the base that diagonalizes $\hat{\sigma}_i\otimes\hat{\sigma}_j$.
    After each unit a PBS separates H and V contributions and by correlating the outcomes of first and second unit it is possible to reconstruct $\langle\hat{\sigma}_i\otimes\hat{\sigma}_j\rangle$}.
    \label{fig:tomosetup}
\end{figure}
Fig. \ref{fig:tomosetup} shows a sketch of the tomography setup.
Photons coming from the TM setup travel to a 50:50 BS, after which, two combinations of a half- and quarter-waveplates installed into motorized rotation mounts implement projections on the Pauli bases according angle settings:\\
\begin{center}
    \begin{tabular}{c|c|c|}
            & HWP & QWP\\
           \hline
           $\hat{\sigma_{x}}$ & 22.5° & 0°\\
           $\hat{\sigma_{y}}$ & 45° & -45°\\
           $\hat{\sigma_{z}}$ & 0° & 0°
    \end{tabular}
\end{center}
Each combination implements a tomographic unitary $\hat U_i$, with $i=x,y,z$, which rotates the local polarization from the basis that diagonalizes $\hat{\sigma}_i$ to the H-V one. 
This allows to implement the projective measurement in the selected basis, using a PBSs to separate H and V polarized light and direct it to two SNSPDs.\\ 
The expectation values $\langle\hat{\sigma}_i\otimes\hat{\sigma}_j\rangle$ are obtained using many copies of $\hat{\rho}$ and measuring coincidences between detector pairs: $HH'$, $HV'$, $VH'$ and $VV'$, whereby:
\begin{equation}\label{eq:correlation}
    \langle\hat{\sigma}_i\otimes\hat{\sigma}_j\rangle=\frac{C^{(i,j)}_{HH'}-C^{(i,j)}_{HV'}-C^{(i,j)}_{VH'}+C^{(i,j)}_{VV'}}{C^{(i,j)}_{HH'}+C^{(i,j)}_{HV'}+C^{(i,j)}_{VH'}+C^{(i,j)}_{VV'}},
\end{equation} 
where the superscripts indicate the combination of unitaries applied to the photons.
A complete reconstruction of $\hat{\rho}$ requires measuring expectation values of the form: $\langle\mathbb{I}\otimes\hat{\sigma}_j\rangle$, $\langle\hat{\sigma}_i\otimes\mathbb{I}\rangle$ and $\langle\mathbb{I}\otimes\mathbb{I}\rangle$, where $\mathbb{I}$ is the $2\times2$ identity.
While the latter may be obtained as a normalization form all the others, the former two can be still obtained as the $\langle\hat{\sigma}_i\otimes\hat{\sigma}_j\rangle$ combining coincidences as follows:
\begin{equation}\label{eq:correlation}
    \langle\hat{\sigma}_i\otimes\mathbb{I}\rangle=\frac{C^{(i,i)}_{HH'}+C^{(i,i)}_{HV'}-C^{(i,i)}_{VH'}-C^{(i,i)}_{VV'}}{C^{(i,i)}_{HH'}+C^{(i,i)}_{HV'}+C^{(i,i)}_{VH'}+C^{(i,i)}_{VV'}},
\end{equation}     
\begin{equation}\label{eq:correlation}
    \langle\mathbb{I}\otimes\hat{\sigma}_j\rangle=\frac{C^{(j,j)}_{HH'}-C^{(j,j)}_{HV'}+C^{(j,j)}_{VH'}-C^{(j,j)}_{VV'}}{C^{(j,j)}_{HH'}+C^{(j,j)}_{HV'}+C^{(j,j)}_{VH'}+C^{(j,j)}_{VV'}}.
\end{equation}     
The density operator $\hat{\rho}$ is obtained considering that a generic two-qubit state may be written as $\hat{\rho}=\sum_{ij}s_{i,j}\hat{\sigma}_i\otimes\hat{\sigma}_j$, with $i=0,x,y,z$ and $\hat{\sigma}_0=\mathbb{I}$.\\
To be able to reconstruct $\hat{\rho}$ it is necessary to perform a sufficient amount of measurements, $i.e.:$ set enough combinations of $\hat U_i$ and $\hat U_j$.
\begin{figure}[h!]
    \centering
    \includegraphics[width=1.001\columnwidth]{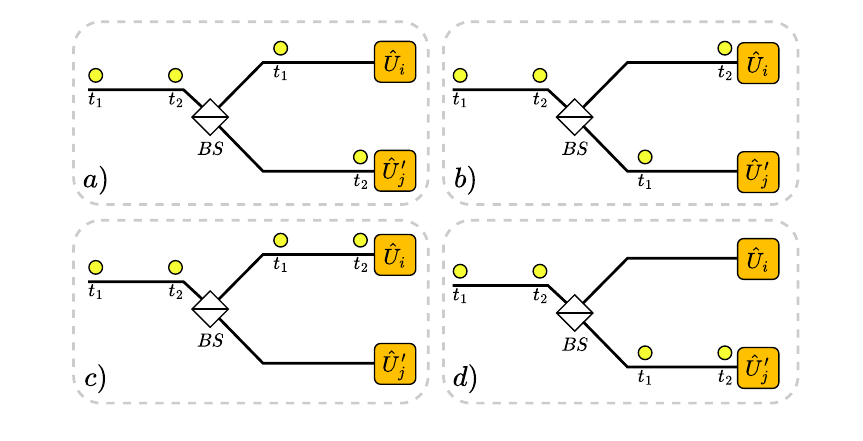}
    \caption{Starting with two photons in two \tbs$\ t_1$ and $t_2$, the four outcomes a)-b) after the BS are possible. 
    The two \tbs$\ $are either separated (cases a) and b)) or sent to the same detection unit. 
    Depending on the specific outcome the state is projected on the eigenbasis of $\hat{\sigma}_i\otimes\hat{\sigma}_j$ , $\hat{\sigma}_j\otimes\hat{\sigma}_i$, $\hat{\sigma}_i\otimes\hat{\sigma}_i$ and $\hat{\sigma}_j\otimes\hat{\sigma}_j$.}
    \label{fig:time_tomo}
\end{figure}
Considering a case where the two photons encoding the two qubit state $\hat{\rho}$ occupy two \tbs$\ t_1$ and $t_2$, upon reaching the 50:50 BS, the four outcomes shown in Fig. \ref{fig:time_tomo} are possible. 
Accordingly, the two photons are either separated (cases a) and b)) or directed to the same tomographic unit (cases c) and d)). 
In cases a) and b) the state is projected on the eigenbasis of $\hat{\sigma}_i\otimes\hat{\sigma}_j$ and $\hat{\sigma}_j\otimes\hat{\sigma}_i$, while in the two remaining cases $\hat{\rho}$ is projected on the eigenbasis of $\hat{\sigma}_i\otimes\hat{\sigma}_i$ and $\hat{\sigma}_j\otimes\hat{\sigma}_j$.
Therefore, if $t_1\neq t_2$, the three settings: ($\hat U_z,\hat U_x$), ($\hat U_z,\hat U_y$) and ($\hat U_x,\hat U_y$) are sufficient to reconstruct the state postselecting on events matching the four possible outcomes.\\
It is possible to reconstruct $\hat{\rho}$ even when $t_1=t_2$, however, in this case all outcomes where the two photons are not separated become unusable as our detectors do not have photon number resolution.
For this reason to reconstruct the state the settings: ($\hat U_z,\hat U_z$), ($\hat U_x,\hat U_x$) and ($\hat U_y,\hat U_y$) must be added to the ones seen for the previous case.
This allows to reconstruct the state by postselecting only on outcomes where photons were separated by the 50:50 BS.\\        
\clearpage{}
\bibliographystyle{apsrev4-2}
\bibliography{biblio}
\end{document}